\documentclass[a4paper, 11pt, onecolumn, accepted=2024-12-19]{quantumarticle}
\pdfoutput=1
\usepackage[table]{xcolor}
\usepackage{amsfonts,amsmath,amsthm,amssymb,braket,enumerate,enumitem,xcolor,float, bbm,mathtools}
\usepackage{authblk}

\usepackage{thmtools}
\usepackage[framemethod=tikz]{mdframed}

\usepackage{pgf}
\usepackage[position=bottom]{subcaption}

\newcounter{algorithm}
\newenvironment{algorithm}[1]
{
    \begin{mdframed}[
        leftline=false,
        rightline=false,
    ]
    \refstepcounter{algorithm}
    \textbf{Algorithm~\thealgorithm:} \quad #1
    \begin{enumerate}
}
{
    \end{enumerate}
    \end{mdframed}
}

\usepackage{thmtools}
\usepackage{thm-restate}

\makeatletter
\renewcommand*\env@matrix[1][*\c@MaxMatrixCols c]{%
  \hskip -\arraycolsep
  \let\@ifnextchar\new@ifnextchar
  \array{#1}}
\makeatother

\newcommand{\C}{\mathbb{C}}
\newcommand{\I}{\mathbb{I}}

\newcommand{\Z}{\mathbb{Z}}
\newcommand{\cA}{\mathcal{A}}
\newcommand{\cC}{\mathcal{C}}

\newcommand{\cP}{\mathcal{P}}
\newcommand{\cS}{\mathcal{S}}
\newcommand{\cU}{\mathcal{U}}

\newcommand{\Stim}{\texttt{Stim \!}}

\newcommand{\Sone}{[S\textsubscript{V}]}
\newcommand{\Stwo}{[S\textsubscript{Q}]}
\newcommand{\Sthree}{[S\textsubscript{P}]}

\newcommand{\Cone}{[C\textsubscript{U}]}
\newcommand{\Ctwo}{[C\textsubscript{T}]}

\newcommand{\kz}{\ket{\vec{z\,}}}

\newtheorem{theorem}{Theorem}
\newtheorem{defn}{Definition}

\newtheorem{lemma}{Lemma}

\theoremstyle{remark}

\newcommand{\supp}{\ensuremath{\text{supp}}}

\usepackage{doi}
\usepackage[url=false]{biblatex}
\usepackage{xurl}
\author[1]{Nadish de Silva}
\email{nadish\_de\_silva@sfu.ca}
\author[2]{Wilfred Salmon}
\author[1]{Ming Yin}

\affil[1]{Department of Mathematics, Simon Fraser University}
\affil[2]{Department of Applied Mathematics and Theoretical Physics, University of Cambridge}

\title{Fast algorithms for classical specifications of stabiliser states and Clifford gates}

\begin{document}

\maketitle

\begin{abstract}
The stabiliser formalism plays a central role in quantum computing, error correction, and fault tolerance.  Conversions between and verifications of different specifications of stabiliser states and Clifford gates are important components of many classical algorithms in quantum information, e.g.\ for gate synthesis, circuit optimisation, and simulating quantum circuits.  These core functions are also used in the numerical experiments critical to formulating and testing mathematical conjectures on the stabiliser formalism.  

We develop novel mathematical insights concerning stabiliser states and Clifford gates that significantly clarify their descriptions.  We then utilise these to provide ten new fast algorithms which offer asymptotic advantages over any existing implementations.  We show how to rapidly verify that a vector is a stabiliser state, and interconvert between its specification as amplitudes, a quadratic form, and a check matrix.  These methods are leveraged to rapidly check if a given unitary matrix is a Clifford gate and to interconvert between the matrix of a Clifford gate and its compact specification as a stabiliser tableau.    

For example, we extract the stabiliser tableau of a $2^n \times 2^n$ matrix, promised to be a Clifford gate, in $O(n 2^n)$ time.  Remarkably, it is not necessary to read all the elements of a Clifford gate matrix to extract its stabiliser tableau.  This is an asymptotic speedup over the best-known method that is exponential in the number of qubits.

We provide implementations of our algorithms in \texttt{Python} and \texttt{C++} that exhibit vastly improved practical performance over existing algorithms in the cases where they exist.
\end{abstract}

\newpage
\tableofcontents
\newpage

\section{Introduction}
The stabiliser formalism is an essential pillar of the theory of quantum computation. It consists of a particularly natural subset of quantum states, gates, and measurements: stabiliser states, Clifford gates, and Pauli measurements, respectively (see Section \ref{background} for precise definitions).  The stabiliser formalism has played a key role in quantum information from its earliest days.  For example, communication protocols such as Wiesner's quantum money scheme \cite{wiesner1983conjugate}, BB84 key distribution \cite{bennett2014quantum}, superdense coding \cite{bennett1992communication}, quantum teleportation \cite{bennett1993teleporting}, and entanglement distillation \cite{bennett1996concentrating} all utilise elements of the stabiliser formalism.  The first proofs of nonlocality, an important topic in quantum foundations, used stabiliser states such as the Bell state \cite{bell1964einstein} and the GHZ state \cite{greenberger1990bell}.  Early quantum algorithms such as the Deutsch-Jozsa \cite{deutsch1992rapid}, Bernstein-Vazirani \cite{bernstein1993quantum}, and Simon algorithm \cite{simon1997power} use only stabiliser operations on their input (given as an oracle).

The elements of the stabiliser formalism were first collectively defined in the context of quantum error-correction \cite{gottthesis}.  Nearly all quantum error-correcting codes presently under study are examples of \emph{stabiliser codes}.  These codes use stabiliser states to encode basic quantum data, i.e.\ logical basis states, and Pauli measurements to detect errors.  The group of Clifford gates, which arise as symmetries of the group of Pauli measurements, represent operations that are easily (e.g.\ transversally) performed fault-tolerantly on encoded data \cite{gottesman2010introduction}.  

The Eastin-Knill theorem \cite{eastin2009restrictions} precludes universal fault-tolerant quantum computation using only a transversal set of quantum gates, which significantly complicates fault-tolerant frameworks. To circumvent this, most proposed fault-tolerant schemes use so-called magic states, which lead to universal quantum computation when taken with stabiliser operations \cite{bravyi2012magic}. Thus, most fault-tolerant quantum architectures rely on stabiliser operations and magic state generation, placing the stabiliser formalism at the heart of experimentally-realisable quantum computers.  In particular, considerable attention is paid to leveraging the stabiliser formalism to minimise the use of costly magic states when synthesising gates or circuits. A deeper understanding of these subjects will hasten the arrival of practical quantum computers.

The stabiliser formalism has also played an central role in the study of classical simulation algorithms for quantum circuits since the introduction of the the Gottesman-Knill theorem \cite{gottesman1998heisenberg, aaronson2004improved, anders2006fast}.  The theorem now has many variants \cite{jozsa2013classical}, all of which roughly state that adaptive circuits restricted to stabiliser operations can be classically efficiently simulated. At the core of the Gottesman-Knill theorem is the observation that elements of the stabiliser formalism, as a result of their defining symmetries, have classical descriptions that are much more compact than their standard Hilbert space descriptions.

Extensions of the Gottesman-Knill theorem are used to simulate universal quantum computation: e.g.\ the stabiliser rank algorithm \cite{bravyi2016trading} and the sum-over-Cliffords algorithm \cite{bravyi2019simulation}.  These algorithms reduce the case of simulating arbitrary circuits to the stabiliser case, with overhead scaling with how `nonstabiliser' the circuit and input state are.  Such algorithms have practical importance for testing quantum circuits, as well as theoretical importance in the study of quantum advantage.


Stabiliser operations are used across many other areas within quantum information.  Some examples include resource states for measurement-based quantum computation \cite{raussendorf2000quantum}, randomised benchmarking \cite{knill2008randomized}, and quantum learning algorithms \cite{huang2020predicting}.  There is substantial interest in, and active research on, finding good quantum protocols for the verification of stabiliser states \cite{stab_state_verif} and Clifford gates \cite{gate_verif}.  

Given the importance of the stabiliser formalism to so many areas of quantum information and computation, it is an urgent task to develop fast classical algorithms for working with its elements.  Improvements to the core subroutines for computations using the stabiliser operations will find applications in, for example: classical simulation algorithms, circuit compilation, and gate synthesis.  They will also enable better numerical experiments leading to the development of the mathematical theory of the stabiliser formalism.  

In this article, we give ten novel fast algorithms for working with classical descriptions of elements of stabiliser theory. In particular, we consider three descriptions of $n$-qubit stabiliser states: \\

    \vspace{-5pt}\noindent \Sone\ \quad as a complex \emph{vector} of amplitudes, \\
    
    \vspace{-5pt}\noindent \Stwo\ \quad as a \emph{quadratic} form, a linear map, and  an affine subspace of $\Z_2^n$,\\
    
    \vspace{-5pt}\noindent \Sthree\ \quad as a check matrix, i.e.\ a compact list of \emph{Pauli} gate generators for the stabiliser group.\\

We also consider two representations of $n$-qubit Clifford gates:\\

    \vspace{-5pt}\noindent \Cone\ \quad  as a \emph{unitary} matrix,\\
    
    \vspace{-5pt}\noindent \Ctwo\ \quad  as a list of $2n$ Pauli gates representing the images of basic Pauli gates under conjugation, i.e.\ a \emph{tableau}.\\

We give algorithms for interconverting between these descriptions, as well as algorithms for verifying whether a given state vector is a stabiliser state, or a given unitary is a Clifford gate.  We achieve very large reductions in their runtimes with respect to existing implementations, by developing mathematical insights that conceptually clarify the elements of the stabiliser formalism.

Implementations of our algorithms in \texttt{Python} and \texttt{C++} can be found at \url{https://github.com/ndesilva/stabiliser-tools}.

\begin{itemize}

\item In Subsection \ref{results}, we summarise our novel algorithms and their complexity advantages. 

\item In Subsection \ref{applications}, we suggest possible domains wherein applications of our algorithms can be useful. 

\item In Section \ref{background}, we provide the necessary definitions and give the precise definitions of the various inputs and outputs of our algorithms. We also describe the na\"{i}ve brute force methods for verifying that a vector or matrix is a stabiliser state or a Clifford gate respectively and for interconverting between their descriptions.

\item In Sections \ref{stabsec} and \ref{cliffsec}, we develop novel theory concerning stabiliser states and Clifford gates respectively.  We use these developments to give much faster algorithms for verifying that a vector or matrix is a stabiliser state or a Clifford gate.  We also give much faster algorithms for interconverting between their descriptions.

\item In Section \ref{sec:complexity_analysis}, we give detailed analyses of the worst-case asymptotic time complexity of our ten algorithms.

\item In Appendix \ref{sec:comparison_to_implementations}, we give detailed analyses of the worst-case asymptotic time complexity of existing implementations of the functions we describe, for the cases where they exist.

\item In Appendix \ref{benchmarks}, we provide plots of timed benchmarks of our \texttt{C++} implementations.  Wherever possible, we compare with existing implementations from \texttt{Stim} \cite{gidney2021stim} and \texttt{Qiskit} \cite{Qiskit}.  We find absolute advantages of some orders of magnitude for small numbers of qubits; these advantages are guaranteed to become arbitrarily large as $n$ grows.


\end{itemize}

\subsection{Summary of results}\label{results}

In this work, we describe ten novel algorithms for interconverting between the above descriptions and for verifying that a candidate description is valid.  In the table on the following page, we compare the worst-case asymptotic complexity of our algorithms to the state-of-the-art. The guide to interpreting these tables immediately follows them.

In all cases where algorithms performing the same functionality are known to exist, we find that our algorithm has a complexity advantage of a factor of at least $n$, the number of qubits in the system.  
The complexity of na\"{i}ve brute force methods are analysed in Subsections \ref{stabback} and \ref{cliffback}.  Improvements on brute force are given in \cite{gidney2021stim} and \cite{Qiskit}; see Appendix \ref{sec:comparison_to_implementations} for their complexity analyses.

Given that proposed error-correcting systems require components built of many-qubit subsystems \cite{dalzell2023quantum}, improvements of a factor of $n$ are already highly consequential. In some cases, we find our algorithm has an advantage by a factor of at least $N = 2^n$: an exponential or higher improvement.   


In practice, our algorithms exhibit excellent performance for any number of qubits.  We achieve this by working directly with the mathematical description in question.  See Appendix \ref{benchmarks} for timed benchmarks of our \texttt{C++} implementations against existing algorithms \cite{gidney2021stim, Qiskit} in the cases where they exist.

In two cases, we give conversion algorithms for which no alternative implementations exist.

\newpage
\begin{table}[h]
\def\arraystretch{2}
\centering
\begin{tabular}{|m{1cm}|m{3.5cm}|m{3.5cm}|m{3.5cm}|}
\hline
     & \textbf{\Sone} & \textbf{\Stwo} & \textbf{\Sthree} \\ \hline
\textbf{\Sone} &   $N n^2 \to N $   &    ? $\to N $    & $N n^2 \to N $    \\ \hline
\textbf{\Stwo} &  $N n^2 \to N n$   &   \cellcolor[HTML]{EFEFEF}$n^2$   &  $N n^2 \to n^3$   \\ \hline
\textbf{\Sthree} &   $N^4 / Nn^2 \to N n$   &    ? $\to n^3$ &  \cellcolor[HTML]{EFEFEF}$n^3$    \\ \hline
\end{tabular}
\end{table}

\begin{table}[h]
\def\arraystretch{1.5}
\centering

\begin{tabular}{|m{1cm}|m{5.46cm}|m{5.46cm}|}
\hline
     & \textbf{\Cone} & \textbf{\Ctwo} \\ \hline
\textbf{\Cone} &  $N^2 n^2 \to N^2 n$    &    $N^2 n^2 \to N n$      \\ \hline
\textbf{\Ctwo} &   $N^2 n^2 \to N^2 n$   &   \cellcolor[HTML]{EFEFEF}$n^3$    \\ \hline

\end{tabular}

\vspace{5pt}\caption{Complexity of existing methods vs.\ new methods.}

\end{table}
\nopagebreak
\vspace{5pt} \noindent \begin{minipage}{\textwidth}
\begin{itemize}

\item Grey cells correspond to simple tasks, all related to the verification of compact descriptions, for which there exists an obvious method that is nearly optimally fast.

\item Diagonal entries of tables correspond to algorithms for verifying candidate descriptions.  For example, the (\Sone,\Sone)-entry in the top-left of the table describes the problem of taking as input a vector in $\C^N$ and deciding whether it represents a valid stabiliser state.

Off-diagonal entries of tables correspond to algorithms for converting from one valid description to another.  For example, the (\Sone,\Sthree)-entry in the top-right of the table describes the problem of taking as input a vector in $\C^N$ that represents a valid stabiliser state and gives as output its check matrix.

\item An entry of the form $X \to Y$ indicates that the best currently-known technique requires time $\Omega(X)$ whereas our methods require only  $O(Y)$ time.

An entry of the form $X_1 / X_2 \to Y$ indicates that the best currently-known technique that is guaranteed to succeed requires time $\Omega(X_1)$ and the best currently-known technique that succeeds with high probability requires time $\Omega(X_2)$ whereas our methods require only  $O(Y)$ time.

\item An entry of the form $? \to Y$ indicates that there is neither an existing implementation nor an obvious method for the interconversion task in question; we give a method that requires time $O(Y)$.

\end{itemize}
\end{minipage}

\subsection{Applications}\label{applications}

The need for fast classical algorithms for verifying and converting descriptions of stabiliser elements is evidenced by online queries \cite{stackexchange1, stackexchange2} and their inclusion in popular software packages \cite{gidney2021stim,Qiskit}.  This is no surprise given the ubiquity of the stabiliser formalism in quantum information.  We identify a number of potential applications in the areas of classical simulation of quantum circuits, gate synthesis, and circuit optimisation; we are sure of the existence of other potential domains of applicability that we cannot anticipate.  We conclude by suggesting how our methods can be used to formulate and test mathematical conjectures concerning the stabiliser formalism. 

\paragraph{Classical simulation.} Given the classical efficiency of the Gottesman-Knill algorithm, it is the standard method for simulating the stabiliser formalism. However, in some regimes, alternative simulation techniques can be faster.  Our conversions are useful at the interfaces between these regimes. For example, recent work by de Beaudrap and Herbert \cite{de2022fast} shows that stabiliser simulation can be made faster in certain cases using the \emph{quadratic form expansion}, which is equivalent to what we term a quadratic form triple, of a stabiliser state. These cases include: performing stabiliser operations where the quantum state has a small computational basis expansion; simulating deterministic single-qubit Pauli measurements; and, particularly of interest, simulating local syndrome measurements for encoded stabiliser circuits. 
In contrast, Gottesman-Knill-type algorithms use the check matrix form of a stabiliser state. Thus our conversions to and from the quadratic form are useful for employing a hybrid algorithm that makes use of the strengths of both types of simulation.

Much work has been done on classical simulation of general quantum circuits by separating stabiliser subcircuits from nonstabiliser parts, e.g.\ recently by Smith et al.\ \cite{smith2023clifford}. Since many popular quantum circuit simulators use state vectors (e.g.\ \cite{quantum_ai_team_and_collaborators_2021_5544365}), our conversions from the other specifications of a stabiliser state to the state vector will speed up the classical processing at these stabiliser-nonstabiliser interfaces.

Next, we note that stabiliser operations together with access to magic states are sufficient for universal quantum computation. Our tableau-to-unitary matrix conversion for Clifford gates can prove useful if one wishes to apply a Clifford, perhaps resulting from back-propagation of Pauli measurements, to initial magic states. As another example, in the sum-over-Cliffords method \cite{bravyi2019simulation}, a unitary is decomposed as a linear combination of Clifford gates which are easier to simulate.  If this decomposition is found in unitary form (perhaps by recursively subtracting the closest Clifford gate from a unitary), our unitary matrix-to-tableau conversion would accelerate the algorithm. 

\paragraph{Gate synthesis.} Here we describe an example  application of our methods that has already improved the efficiency of gate synthesis algorithms.  Clifford isometries are a generalisation of standard Clifford gates which are of considerable interest in areas such as magic state distillation \cite{bravyi2005universal}. Recent work by Kliuchnikov, Beverland and Paetznick \cite{kliuchnikov2023stabilizer}, followed by Kliuchnikov and Schonnenbeck \cite{kliuchnikov2024stabilizer}, describes an efficient way to determine whether a unitary (in complex matrix form) is a Clifford isometry, and if so, to compile it over a chosen gateset \cite{de2022graph}. The method uses the Choi stabiliser state of the operator, which must be converted from state vector form to check matrix form before the next steps can be used to find the tableau form of the Clifford isometry \cite{audenaert2005entanglement}. Our method of converting the Choi state to check matrix form is used to improve the efficiency of this method \cite[p.\ 7]{kliuchnikov2024stabilizer}. 

\paragraph{Circuit synthesis and optimisation.} General quantum circuits are commonly compiled in terms of Pauli exponentials \cite{mattpc}, e.g.\ for $T$-count reduction algorithms \cite{zhang2019optimizing} or in the context of quantum chemistry problems such as estimating ground state energies of Hamiltonians \cite{cowtan2020generic}. A Pauli exponential has the general form $R_P(\theta) = e^{-i\frac{\theta}{2} P}$, where $P$ is a Pauli gate, and can be readily diagonalised into a form such as $R_Z(\varphi)$. The action of diagonal Pauli exponentials and CNOT gates \cite{Nam_2018} can be readily expressed in a phase polynomial form which is somewhat more general than the quadratic form triple of a stabiliser state. Phase polynomials have been extensively studied for circuit optimisation \cite{amy2014polynomial}, e.g.\ in order to reduce the count of two-qubit gates. Thus, having fast algorithms for converting to and from the quadratic form representation of stabiliser states may aid the synthesis of quantum circuits, particularly if one wishes to use phase polynomials to perform some kind of optimisation on a subcircuit. On a related note, it may help to use these conversions in cases where one wishes to use a Clifford to conjugate a fixed Pauli to another fixed Pauli \cite{mattpc, Amy:2022ljh}, since it may be useful to perform phase polynomial optimisation on a number of different such Cliffords to decide which one to choose.

\paragraph{Mathematical exploration.} Our methods are also useful as research tools in quantum information by enabling faster and larger numerical experiments.  As an example, the present work was initially motivated by the problem of searching for low-rank stabiliser decompositions \cite{bravyi2016trading} of magic states.  Unlike previous approaches, which search over sets of stabiliser states, our approach is to generate promising small linear combinations of vectors and verify that they are indeed stabiliser states.  As another example, in the context of searching for proofs of quantum advantage via boson sampling \cite{aaronson2011computational}, one method under development \cite{shanepc} involves translating photonics circuits into unitary matrices and checking whether the result is a Clifford gate.  More broadly, our methods can be used to formulate and test mathematical conjectures.  This may take the form of numerical experiments that check whether the outcome of a randomised procedure is always a stabiliser state or Clifford gate.  Such experiments could deploy our algorithms millions or billions of times, multiplying their performance advantages.\\

We have given above possible use cases for all ten of our novel algorithms.  We anticipate that many more will be found by those with expertise in areas beyond our own.

\section{Background}\label{background}

Let $n$ be a positive integer and $N = 2^n$.  We denote the computational basis states of an $n$-qubit system by $\kz \in \C^N$ for $\vec z \in \Z_2^n$.  Given $n$ unitaries $U_1, \dots, U_n$ and a vector $\vec{p}$ of $n$ integers, we denote by $U^{\vec{p}}$ the product $U_1^{p_1} \cdots U_n^{p_n}$.  The basic Pauli gates are $Z_i,X_i$: i.e.\ $Z,X$ respectively on the $i$-th qubit and identity on all others.  Thus, $Z^{\vec p}$ and $X^{\vec q}$, with $\vec p, \vec q \in \Z_2^n$, are defined by \begin{equation}\label{paulidef}Z^{\vec p} \kz = (-1)^{\vec p \cdot \vec z} \kz \quad \quad \quad X^{\vec q}\ket{\vec z} = \ket{\vec z + \vec q} \end{equation} where the addition is entrywise and in $\Z_2$.

\begin{defn}\label{paulidefn}The group of \emph{Pauli gates} is the subgroup of $\cU(2^n)$ generated by the basic Pauli gates and $i \I$: \begin{equation}\cP_n = \{(-1)^c (-i)^d X^{\vec{q\,}}  Z^{\vec{p}} \; | \;  (\vec{p}, \vec{q\,}) \in \Z_2^{2n}, \; c,d \in \Z_2 \}.\end{equation}
\end{defn}

The Pauli gates $X^{\vec{q}_1} Z^{\vec{p}_1}$ and $X^{\vec{q}_2} Z^{\vec{p}_2}$ commute if and only if $[(\vec{p}_1, \vec{q}_1),(\vec{p}_2, \vec{q}_2)] \equiv \vec{p}_1\cdot \vec{q}_2 - \vec{p}_2\cdot \vec{q}_1 = 0$ over $\Z_2$.  A set of Pauli gates is said to be independent if no nontrivial product of them is a multiple of the identity matrix. 

\subsection{Stabiliser states}\label{stabback}

Here, we define stabiliser states and the three ways they can be specified (in the first case, precisely, and in the latter two cases, up to phase).

\begin{defn}\label{stabsbgrpdefn}A \emph{stabiliser subgroup} is a maximal abelian subgroup $\cS$ of $\cP_n$ that does not contain $-\I$.
\end{defn}

The stabiliser subgroups $\cS \subset \cP_n$ are all of size $N$.  A stabiliser subgroup can be specified (nonuniquely) by a \emph{check matrix}: a $n \times (2n+1)$ matrix over $\Z_2$ \begin{equation}\begin{bmatrix}
\vec{q}_1 & \vec{p}_1 & c_1\\
\vdots & \vdots & \vdots\\
\vec{q}_n & \vec{p}_n & c_n\\
\end{bmatrix}\end{equation} that collates a set of $n$ commuting Pauli gates $(-1)^{c_i} (-i)^{\vec{p} \cdot \vec{q}} Z^{\vec{p}} X^{\vec{q\,}}$ that generate $\cS$.

\begin{defn}\label{stabdefn}A \emph{stabiliser state} is a state $\ket{s} \in \C^{N}$ such that: \begin{equation}P\ket{s} = \ket{s}\end{equation} for all $P \in \cS$ in some stabiliser subgroup $\cS \subset \cP_n$.
\end{defn}

A stabiliser state $\ket{s}$ is specified up to phase by a (nonunique) check matrix for the stabiliser subgroup ${\cS_{\ket{s}} = \{P \in \cP_n \;|\; P\ket{s} = \ket{s}\}}$.

\begin{theorem}[Dehaene-De Moor \cite{dehaene2003clifford}, 2003]\label{dehaenedemoor}
Every stabiliser state $\ket{s}$ (up to phase and normalisation) is specified by a triple $(\cA, Q, \ell)$ where $\cA \subset \Z_2^n$ is the affine subspace $V + \vec{z}_0$ for $V \subset \Z_2^{n}$ a vector subspace and $\vec{z}_0 \in \Z_2^{n}$, $Q:V\to\Z_2$ is a quadratic form, and $\ell: V\to\Z_2$ is a linear map: \begin{equation}\label{ddeqn}\ket{s} \propto \sum_{\vec z \in V}  (-1)^{Q(\vec z)} i^{\ell(\vec z)} \ket{\vec z + \vec{z}_0}.\end{equation}
    
\end{theorem}

We will refer to this description as the \emph{quadratic form triple} of a stabiliser state for brevity.  Here, $\cA$ is of dimension $k \leq n$ and can be specified by $k$ basis vectors $\vec z_1, \dots, \vec z_k$ for $V$, and a shift vector $\vec z_0$.

Theorem \ref{dehaenedemoor} has been further generalised to a more general class of states: see Ref. \cite{ni2015non} for details.\\
We have thus described three ways of specifying a stabiliser state (up to phase): \\

    \vspace{-5pt}\noindent \Sone\ \quad  as a complex vector of amplitudes: an element of $\C^N$,\\
    
    \vspace{-5pt}\noindent \Stwo\ \quad  as an affine subspace, a linear map, and a quadratic form: an element of $(\Z_2^{n})^{k+1} \times \Z_2^n \times \Z_2^{n(n+1)}$,\\
    
    \vspace{-5pt}\noindent \Sthree\ \quad  as a check matrix: an element of $(\Z_2^{2n+1})^{n}$.\\

\vspace{-5pt} The brute force test for whether $\ket{\psi}\in\C^N$ is a stabiliser state requires iterating through each of the $N^2$ Pauli gates $P$, computing $P\ket{\psi}$, and determining the number of Pauli gates that stabilise $\ket{\psi}$.  Verifying \Stwo\ requires checking that the proposed basis vectors are linearly independent, taking time $O(n^2)$. Verifying \Sthree\ requires checking that the $n$ Pauli gates are independent and commute, which takes time $O(n^3)$.

The brute force method to convert a description from \Sone\ to \Sthree, is to iterate through the Pauli gates until a stabiliser group is constructed.  To convert a description from \Sthree\ to \Sone, one constructs the projection onto the state by summing all the Pauli gates in its stabiliser group and finding the projector's $+1$-eigenvector.  This is very costly as it involves multiplying and inverting matrices of size $N$.  Converting from \Stwo\ to \Sone\ via brute force involves evaluating a quadratic form to find each amplitude; converting from \Stwo\ to \Sthree\ would then require composing the result with the above \Sthree\ to \Sone\ conversion. Methods of converting \Sone\ or \Sthree\ to \Stwo\ are not immediately obvious as the work of Dehaene-De Moor does not readily admit implementation as an algorithm.

\subsection{Clifford gates}\label{cliffback}

\begin{defn}\label{cliffdefn}The group of \emph{Clifford gates} is the normaliser of the group of Pauli gates as a subgroup of the unitary matrices $\cU(N)$: \begin{equation}\cC_n = \{C \in \cU(N) \;|\; C \cP_n C^* \subseteq \cC_n \}.\end{equation}
\end{defn}

The Clifford gates, up to phase, are in correspondence with  \emph{conjugate tuples} \cite[Definition 3.6]{de2021efficient} of Pauli gates: these are $n$-tuples of pairs of Pauli gates $((U_1,V_1),\dots,(U_n,V_n))$ such that $U_i^2 = V_i^2 = \mathbb{I}$, $U_i V_j = (-1)^{\delta_{ij}} V_j U_i$, $U_i U_j = U_j U_i$, and $V_i V_j = V_j V_i$ \cite[Theorem 3.11]{de2021efficient}.  A Clifford gate $C$, up to phase, corresponds to the conjugate tuple $((CZ_1C^*,CX_1C^*),\dots,(CZ_nC^*,CX_nC^*))$.   Each such Pauli, being of order 2, is of the form $(-1)^c(-i)^{\vec{p}\cdot\vec{q}} Z^{\vec{p}} X^{\vec{q\,}}$ and thus specified by $2n+1$ bits.

The conjugate tuple specification of a Clifford gate is thus vastly more compact ($4n^2 + 2n$ bits vs.\ $N^2$ complex numbers) and can be used more easily to e.g.\ act on stabiliser states in check matrix form or conjugate Pauli gates expressed as vectors in $\Z_2^{2n+1}$.  It is well known in the literature as the \emph{stabiliser tableau} of the Clifford gate.


We have thus described 
two ways of specifying a Clifford gate (up to phase):\\

    \vspace{-5pt}\noindent \Cone\ \quad  as a complex matrix $C$: an element of $\C^{N \times N}$\\
    
    \vspace{-5pt}\noindent \Ctwo\ \quad  as a conjugate tuple of Pauli gates $U_i = CZ_iC^*, V_i = CX_iC^*$: an element of $((\Z_2^{2n+1})^{2})^n$.\\

\vspace{-5pt} The brute force method to verify that a matrix $M$ is a Clifford gate is to conjugate the basic Pauli gates and verify that the results are also Pauli gates; this requires conjugating $2N$ matrices of size $N$; every matrix multiplication requires roughly $N^3$ steps.  To verify \Ctwo, one must check that the given Pauli gates satisfy the canonical commutation relations, which takes time $O(n^3)$.

The brute force method to convert a description from \Cone\ to \Ctwo\ is to conjugate the basic Pauli gates as above.  The converse direction is described in \cite[Theorem 3.9]{de2021efficient}; there, it requires a conversion between descriptions \Sone\ to \Sthree\ of a stabiliser state.  We can therefore use our new methods to improve the conversion \Ctwo\ to \Cone, as detailed below.

\section{Stabiliser state algorithms}\label{stabsec}

Here, we describe a rapid method for extracting from a vector of amplitudes of a stabiliser state its quadratic form triple: that is, the conversion \Sone\ $\to$ \Stwo.  We then provide a rapid method for the converse conversion \Stwo\ $\to$ \Sone.  With some further checks, our algorithms also give a natural method for verifying that a complex vector is a stabiliser state. We also show that the conversions between a check matrix and a quadratic form triple and vice versa---\Sthree\ $\to$ \Stwo\ and \Stwo\ $\to$ \Sthree---can be done very quickly.

\subsection[Amplitudes to a quadratic form triple]{\Sone\ $\to$ \Stwo: Amplitudes to a quadratic form triple}

Theorem \ref{dehaenedemoor} already imposes strong constraints on the amplitudes $\braket{\vec z\, | s}$ of a stabiliser state $\ket{s}$: up to a common factor, they must take values in $\{\pm 1, \pm i\}$ and the support $\supp(\ket{s}) = \{\vec z \in \Z_2^n \;|\; \braket{\vec z\, | s} \neq 0\}$ of $\ket{s}$ is an affine subspace $\cA = V + \vec{z}_0$ of dimension $k \leq n$. We can break converting the amplitudes of $\ket{s}$ to a quadratic form triple into two stages:

\begin{enumerate}
    \item Find a basis $\vec{z}_1, \dots, \vec{z}_k$ for $V$ and $\vec{z}_0$ such that the support $\supp(\ket{s}) = \cA = V + \vec{z}_0$.
    \item Find the coefficients of the quadratic form $Q$ and the linear map $\ell$  as in Theorem \ref{dehaenedemoor}.
\end{enumerate}

To perform the first step, we convert $\supp(\ket{s})$ into a vector space by taking $\vec z_0$ to be the first element (in the natural ordering on bitstrings) of $\supp(\ket{s})$ and subtracting $\vec z_0$ from every element of $\supp(\ket{s})$.  Using the following lemma, we can conclude that this procedure leaves us with a \emph{sorted} list of the vectors in $V$.

\begin{lemma}\label{sortedvspace}
    Let $\vec v,\vec w, \vec z_0 \in \Z_2^n$.  Suppose that $\vec v < \vec w$ and $\vec w + \vec z_0 < \vec v + \vec z_0$.  Then $(\vec v + \vec w) + \vec z_0 < \vec z_0$.
\end{lemma}

\begin{proof}
Here, the notation $(\vec a, \vec b)$ denotes the concatenation of bitstrings, i.e.\ a direct sum of vectors.  

As $\vec v < \vec w$, we have that $\vec v = (\vec{v}', 0, \; \vec{c}\,)$ and $\vec w = (\vec{w}', 1, \vec{c}\,)$ for some (possibly empty) substrings $\vec{v}',\vec{w}',\vec{c}$.

Expressing $\vec z_0$ as $\vec z_0 = (\vec d, e, \vec f\,)$, we have $\vec v + \vec z_0 = (\vec{v}' + \vec{d}, e, \vec c  + \vec f\,)$ and $\vec w + \vec z_0 = (\vec{w}' + \vec d, e + 1, \vec c + \vec f\,)$.  Therefore, $\vec v + \vec z_0 > \vec w + \vec z_0$ if and only if $e = 1$.  

If $e=1$, then $(\vec v + \vec w) + \vec z_0 = (\vec{v}' + \vec{w}' + \vec d, 0, \vec f\,) < \vec z_0$.
\end{proof}

Since $\vec z_0$ is the minimal element of $\cA$, we can conclude that the process of subtracting $\vec z_0$ from every element of a sorted list of $\cA$ results in a sorted list.

As we now have $V$ as a sorted list, the basis vectors $\vec{z}_i$ are simply those whose position in the list is a power of 2.  This is a consequence of the following lemma.

\begin{lemma}\label{orderedvectorbasis}
    For any $m \in \mathbb{N}$ and any $\vec x = (x_{m},\dots,x_1) \in \Z_2^m$, denote by $I(\vec x)$ its corresponding integer $\sum_{i=1}^{m} x_i 2^{i-1}$.
    Suppose $V \subseteq \Z_2^n$ is a subspace of dimension $k$ ordered such that $\vec v < \vec v\,' \iff I(\vec v) < I(\vec v\,'$\!) and, in this order, $V = \{\vec{v_0}, \dots, \vec v_{2^k-1}\}$.  Then the map $T: \Z_2^k \to V$ given by $T(\vec x) = \vec v_{I(\vec x)}$ is a linear isomorphism.

\end{lemma}

\begin{proof}

Let $L: V \to \{0,\dots,n\}$ return the position of the leftmost digit, counting from the right end; $L(\vec{v}_0) = 0$.  This function is clearly nondecreasing.

For $j \in \{1,\dots,k\}$: $\vec w_j = \vec v_{2^{j-1}}$ and $p_j = L(\vec w_j)$; $p_0 = 0$.  

For $j \in \{0,\dots,k\}$: define $V_j = \{\vec v_0, \dots, \vec v_{2^j - 1} \} = \{ \vec v \in V \;|\; \vec v < \vec w_{j+1} \}$.

We will prove the following claim by induction: $L^{-1}( \{p_0, \dots, p_j\} ) = V_j$ for $j \in \{0,\dots,k\}$.

This trivially holds for $j = 0$; assume it holds for $j$.  Using the definition of $V_j$ and the induction hypothesis to establish two containments respectively, we see that:
$L^{-1}(p_{j+1}) = V_j + \vec w_{j+1}$.
This set contains $2^j$ distinct, consecutive elements, including $\vec w_{j+1}$, and is disjoint from $V_j$.  It is thus equal to $V_{j+1} \setminus V_j$ and our claim is proved.

Thus, $\{p_k, \dots, p_1\}$ are distinct and so $(\vec w_k, \dots, \vec w_1)$ is an ordered basis of $V$.

Let $T: \Z_2^k \to V$ be given by $T(\vec{x}) = \sum_{i=1}^k x_i \vec w_i$.  We can prove that $\sum_{i=1}^k x_i \vec w_i = \vec v_{I(\vec x)}$ for each $V_j$ by induction on $j$.  This requires the fact that adding $\vec w_{j+1}$ to elements of $V_j$ is an order-preserving operation.  Suppose that $\vec v < \vec v' \in V_j$.  Then $\vec v' + \vec w_{j+1} < \vec v + \vec w_{j+1}$ only if $\vec w_{j+1}$ contains a 1 at the rightmost position in which  $\vec v$ and $\vec v'$ differ.  In this case, $\vec v + \vec v' + \vec w_{j+1} \in V_{j+1} \setminus V_j$ but is strictly less than $\vec w_{j+1}$; a contradiction.
\end{proof}

Having extracted $\vec z_0$ and transformed the support such that  $\supp(\ket{s}) = V = \{\sum_i \alpha_i \vec{z}_i \mid \alpha_i \in \Z_2\}$, we can determine $Q, \ell$ as functions of the vectors $\vec \alpha$.  From the amplitudes corresponding to members of the support with $\vec \alpha$ having Hamming weight 1, we can extract $\ell$.  Similarly, from the amplitudes corresponding to members of the support with $\vec \alpha$ having Hamming weight 1 or 2, we can extract $Q$.  Each such $\vec \alpha$ gives an equation: \begin{equation}\label{amplinsys}
\sum_{1\leq i \leq j \leq k} c_{ij} \alpha_i \alpha_j = Q\left(\sum_i \alpha_i \vec{z}_i\right).\end{equation}  We simply need to solve this inhomogenous linear system of $k(k+1)/2$ equations for the coefficients $c_{ij}$.

\begin{algorithm}
{Convert the amplitudes of a stabiliser state to a quadratic form triple, i.e.\ \Sone\ $\to$ \Stwo}
\label{s1tos2}

\item Choose $\vec z_0$ to be the binary label for the first nonzero amplitude.  Subtract $\vec z_0$ from each element of $\supp(\ket{s}) = V + \vec z_0$ to transform it to the vector space $V$.

\item Sort the binary labels of $\supp(\ket{s}) = V$.  Taking those labels whose position is a power of 2 gives a basis $\{\vec z_i\}$ for this vector space.

\item Extract $\ell$ by inspecting the amplitudes corresponding to $\vec z_i$ and observing whether they are real or imaginary. 

\item Extract $Q$ by solving the linear system \eqref{amplinsys}.

\end{algorithm} 

\subsection[Quadratic form triple to amplitudes]{\Stwo\ $\to$ \Sone: Quadratic form triple to amplitudes}

The obvious conversion method is to evaluate the quadratic form for each element of the support of the state vector.  Indeed, it is surprising that one can improve upon this method.  We obtain an asymptotic advantage by carefully choosing the order in which we iterate through the support.  With this order, we save time by avoiding a fresh evaluation of the quadratic form; instead we update the result of the previous evaluation.

To be precise, let a vector $\vec{\alpha} \in \Z_2^k$ represent the element of the support: $\vec{v}(\vec{\alpha}) = \sum_i \alpha_i \vec{z}_i$. Then, the nonzero elements of the state vector of the stabiliser state are given by $(-1)^{Q(\vec{\alpha})}i^{\ell\vec{\alpha})}\ket{\vec{v}(\vec{\alpha})+\vec z_0}$. Thus, to construct the state vector, we iterate through $\Z_2^k$, setting the nonzero elements of the state vector.

Suppose that one iterates through $\Z_2^k$ in some order $\vec\alpha_0,\dots \vec\alpha_{2^k-1} \in \Z_2^k$. Instead of calculating $\vec{v}(\vec\alpha_i)+\vec z_0$ for every $i$, from scratch, we can keep track of the `current' $\vec\beta_i = \vec{v}(\vec\alpha_i)+\vec z_0$. Then, by linearity,
\begin{equation}
    \vec\beta_{i+1} = \vec \beta_i + \vec v (\vec \alpha_i + \vec \alpha _ {i+1}). \label{eqn:support_update}
\end{equation}

In particular, we iterate through Gray code \cite{Bitner1976gray} (i.e.\ $\vec \alpha_i$ is the $i$-th codeword of the Gray code). This has the useful property that $\vec \alpha_i + \vec \alpha _ {i+1}$ always has Hamming weight 1, and therefore we only need to add a single basis vector to $\vec \beta_i$ to construct $\vec \beta _{i+1}$. Similarly, for $\vec e_i \in \Z_2^k$, the $i$-th standard basis vector,
\begin{align}
    \ell(\vec \alpha + \vec e_i) + \ell(\vec \alpha) &= \ell(\vec e_i), \\
    Q(\vec \alpha + \vec e_i) + Q(\vec \alpha) &= \sum_{j < i} (Q_{ij} + Q_{ji}) \vec \alpha_j. \label{eqn:phase_update} 
\end{align}
Thus, if we keep track of the overall phase $(-1)^{Q(\vec{\alpha})}i^{\ell\vec{\alpha})}$, it takes time linear in $n$ to perform each update as we iterate through the Gray code. 

\begin{algorithm}
{Convert a quadratic form triple to the amplitudes of a stabiliser state, i.e.\ \Stwo\ $\to$ \Sone}
\label{s2tos1}

\item Initialise an array of complex numbers $\Psi\in\C^N$, meant to be our state vector, to be all zeroes. Set $\vec \beta_0 = \vec z _0$ and $t_0 = 1$.

\item Let $\vec\alpha_i$ be the $i$-th codeword of the Gray code (on $k$ bits). Iterate through the Gray code, starting with $i=1$ (recall that the first codeword has $i=0$). Supposing that $\vec\alpha_{i} + \vec\alpha_{i-1} = \vec e_j$ for some $j\in\Z_k$, use Equations \eqref{eqn:support_update}--\eqref{eqn:phase_update} to find $\vec \beta _i$ and $t_i$. Set $\Psi[\beta_i] = t_i$.

\end{algorithm} 

\subsection[Verifying a vector is a stabiliser state]{\Sone: Verifying a vector is a stabiliser state}\label{sec:verif_stab_state}

One can extend Algorithm \ref{s1tos2} to verify that a given vector is a stabiliser state. First, we run Algorithm \ref{s1tos2}, and reject if any of the steps of the algorithm fail. Second, we must check whether the support of the state has size $2^k$. Finally, we must also check that the remaining nonzero entries of the state vector are consistent with $\ell$ and $Q$. To do this, one runs Algorithm \ref{s2tos1} (\Stwo\ $\to$ \Sone), and checks that the resulting state vector is equal to the input.

\subsection[Check matrix to a quadratic form triple]{\Sthree\ $\to$ \Stwo: Check matrix to a quadratic form triple}

We will show that with minimal calculations, one can effectively read off the quadratic form triple data of a stabiliser state from its check matrix.  We begin by establishing some required relations between a stabiliser state expressed as a quadratic form triple and the Pauli gates that stabilise it.  
Recall the expression of the amplitudes of a stabiliser state from Theorem \ref{dehaenedemoor}.  An arbitrary stabiliser state $\ket{s}$ can be expressed in the form: \begin{equation}
    \ket{s} \propto \sum_{\vec z \in V} (-1)^{Q(\vec z)} i^{\ell(\vec z)}  \ket{\vec z + \vec z_0}  \tag{\ref{ddeqn}}
\end{equation} where $V \subset \Z_2^n$ is a vector subspace of dimension $k \leq n$, $\vec z_0 \in \Z_2^n$, and $Q, \ell: V \to \Z_2$ .

Suppose $P = (-1)^c (-i)^{\vec p \cdot \vec q} X^{\vec q} Z^{\vec p}$ stabilises $\ket{s}$: $P\ket{s} = \ket{s}$.  We can see immediately that $\vec q \in V$ or else $P$ changes the support of $\ket{s}$.  Moreover, by comparing the amplitudes of  $\ket{s}$ and $P\ket{s}$, we find that \begin{equation}\label{imgcondn}\ell(\vec{q\,}) = \vec p \cdot \vec q\end{equation}\begin{equation}\label{stabcondn}
    \text{For all }\vec z \in V: \quad Q(\vec z) = c + Q(\vec z + \vec q) + \vec p \cdot (\vec z + \vec q + \vec z_0) + (\vec p \cdot \vec{q\,}) \ell(\vec z), \text{ over } \Z_2.
\end{equation}

Choose $B$ to be a $\Z_2$-bilinear form such that $Q(\vec z) = B(
\vec z, \vec z)$; for example if $Q(\vec z) = \vec z^{\,T} \tilde Q \vec z$ for the upper triangular matrix $\tilde Q$, then we can take $B(
\vec z, \vec z{\,'}) = \vec z^{\,T} \tilde Q \, \vec z{\,'}$.  We can then express Equation \eqref{stabcondn} as \begin{equation}
    [B(\vec q, \vec z) + \prescript{\text{T}\!}{}B(\vec q, \vec z) + (\vec p \cdot \vec{q\,}) \ell(\vec z) + \vec p \cdot \vec z] + [B(\vec q, \vec q) + \vec p \cdot (\vec q + \vec z_0) + c]= 0
\end{equation} where the first term is a linear function of $\vec z \in V$ and the second term is a constant.  This can hold only if both terms are identically zero.  Therefore:
\begin{equation}\label{ceqn}
c = \vec p \cdot (\vec q + \vec z_0) + B(\vec q, \vec q)
\end{equation}\begin{equation}\label{pxeqn}
\vec p \cdot \vec z = B(\vec q, \vec z) + \prescript{\text{T}\!}{}B(\vec q, \vec z) + (\vec p \cdot \vec q) \ell(\vec z)
\end{equation} where the latter equality is as linear functions from $V$ to $\Z_2$.  For each $\vec q \in V$, there is precisely one choice of $c$ and one choice of $\vec p$, up to adding an element of $V^\perp$, such that $P\ket{s} = \ket{s}$.

Now, suppose we are given a check matrix of a stabiliser state.  Without loss of generality, we may perform row reduction and assume that the first $2n$ columns are in reduced echelon form:
\begin{equation}\begin{bmatrix}\label{reducedcheck}
\vec{q}_1 & \vec{p}_1 & c_1\\
\vdots & \vdots & \vdots\\
\vec{q}_k & \vec{p}_k & c_k\\
\vec{0} & \vec{\rho}_1 & \gamma_1\\
\vdots & \vdots & \vdots\\
\vec{0} & \vec \rho_{n-k} & \gamma_{n-k}\\
\end{bmatrix}\end{equation}

We can immediately take the $\vec q_i$ as a basis for $V$.  The shift $\vec z_0$ is any solution to the system of $n-k$ equations $\vec{\rho}_i \cdot \vec z_0 = \gamma_i$.  By Equation \eqref{imgcondn} and the fact that all the stabilising Pauli gates must have order 2, we can extract $\ell$ and define it on the basis for $V$ via the equations $\ell(\vec q_i) = \vec p_i \cdot \vec q_i$.

We can extract the diagonal entries of $\tilde Q$ by rearranging Equation \eqref{ceqn}: \begin{equation}\label{diagq}
\tilde Q_{ii} = B(\vec q_i, \vec q_i) = c_i + \vec p_i \cdot (\vec q_i + \vec z_0).
\end{equation}  
We can extract the strictly upper triangular entries of $\tilde Q$, i.e.\  $\tilde Q_{ij}$ for $1 \leq i < j \leq k$, by rearranging Equation \eqref{pxeqn}:\begin{equation}\label{offdiags}
\tilde Q_{ij} = B(\vec q_i, \vec q_j) = B(\vec q_i, \vec q_j) + \prescript{\text{T}\!}{}B(\vec q_i, \vec q_j) = \vec p_i \cdot \vec q_j + (\vec p_i \cdot \vec q_i)(\vec p_j \cdot \vec q_j).
\end{equation}

\begin{algorithm}
{Convert the check matrix of a stabiliser state to a quadratic form triple, i.e.\ \Sthree\ $\to$ \Stwo}
\label{s3tos2}

\item Row reduce the first two columns whilst updating the sign bits consistently to ensure that the resulting check matrix represents the same stabiliser state up to phase.  Label the result as in Equation \eqref{reducedcheck}.

\item Take the $\vec q_i$ as a basis for $V$ and a solution to $\vec{\rho}_i \cdot \vec z_0 = \gamma_i$ for the shift $\vec z_0$. 

\item Compute $\ell(\vec q_i) = \vec q_i \cdot \vec p_i$ and store these as a vector representing $\ell$ in the ordered basis $\{\vec q_1,\dots,\vec q_k\}$ of $V$.

\item Compute a matrix representing $Q$ relative to the same ordered basis using Equations \eqref{diagq} and \eqref{offdiags}.

\end{algorithm}

\subsection[Quadratic form triple to a check matrix]{\Stwo\ $\to$ \Sthree: Quadratic form triple to a check matrix}

We have shown above that it is relatively straightforward to read off the quadratic form triple data from a check matrix.  Here, we reverse this process. Suppose we are given a basis $\{\vec v_1,\dots,\vec v_k\}$ for $V \subset \Z_2^n$, a shift vector $\vec z_0 \in \Z_2^n$, a $k \times k$ upper triangular matrix $\tilde Q$ over $\Z_2$, and a vector $\ell \in \Z_2^k$.

As in Equation \eqref{reducedcheck}, we can generate the $\vec q_i$ by row reducing the basis $\vec v_i$; we store the resulting change of basis matrix of $\Z_2^k$ and use it to update $\tilde Q,\ell$.  The updated $\tilde Q$ may no longer be upper triangular but can made so by taking $\tilde Q + \tilde Q_l + (\tilde Q_l)^\text{T}$ where $\tilde Q_l$  is the strictly lower triangular part of $\tilde Q$.   

Applying Equation \eqref{pxeqn} to $(\vec p, \vec q) = (\vec{\rho}_i, \vec 0)$, we see that the $\vec{\rho}_i$ can be taken to be a basis of the null space of the matrix whose rows are $\vec q_i$, i.e.\ of $V^\perp$.  We can then extract $\gamma_i$ using $\gamma_i = \vec \rho_i \cdot \vec z_0$.

Using Equations \eqref{imgcondn} and \eqref{pxeqn} we can use our given $\tilde Q$ and $\ell$ to give linear systems for each $\vec p_i$:\begin{equation}\label{pilinsys}
\vec p_i \cdot \vec q_j = \tilde Q_{ij} + \ell(\vec q_i) \ell(\vec q_j).
\end{equation}
These determine each $\vec p_i$ in the basis $\{\vec q_1,\dots,\vec q_k\}$ up to an element of $V^\perp$.

Finally, we compute the $c_i$ using Equation \eqref{ceqn}:\begin{equation}
    c_i = \vec p_i \cdot (\vec q_i + \vec z_0) + \tilde Q_{ii}.
\end{equation}

\begin{algorithm}
{Convert the quadratic form triple of a stabiliser state to a check matrix, i.e.\ \Stwo\ $\to$ \Sthree}
\label{s2tos3}

\item Row reduce the basis $\{\vec v_1, \dots, \vec v_k\}$ of $V$ to find $\{\vec q_1, \dots, \vec q_k\}$ and a basis $\{\vec \rho_1, \dots,\vec \rho_{n-k}\}$ of $V^\perp$.

\item Update $\tilde Q, \ell$ using the change of basis matrix of the previous Step.  Convert $\tilde Q$ to an upper triangular matrix representing the same quadratic form.

\item Compute $\gamma_i = \vec \rho_i \cdot \vec z_0$.

\item Find each of the $\vec p_i$ by solving the underdetermined linear systems of Equation \eqref{pilinsys}.

\item Compute $c_i = \vec p_i \cdot (\vec q_i + \vec z_0) + \tilde Q_{ii}$.

\end{algorithm}

\section{Clifford gate algorithms}\label{cliffsec}

In this section, we describe a highly efficient algorithm for extracting from a Clifford gate $C$, given by its matrix elements, its conjugate tuple: the $2n$ Pauli gates $U_i = CZ_iC^*, V_i = CX_iC^*$.  It can also be used to check if a given unitary matrix $U$ is a Clifford gate.  It works by building upon our above algorithm for extracting generators for the stabiliser group of a stabiliser vector.

\subsection[Matrix entries to a conjugate tuple]{\Cone\ $\to$ \Ctwo: Matrix entries to a conjugate tuple}

The conjugate tuple $U_i = CZ_iC^*, V_i = CX_iC^*$, expressed as $2n$ elements of $\Z_2^{n+1}$, of a Clifford gate can be found in two steps.

We begin by extracting the $U_i$ by examining the stabiliser group of the first column of $C$.  To justify this calculation, we require some lemmata.

\begin{lemma}For a Clifford gate $C$, the stabiliser group of $C\ket{\vec 0}$ is generated by $U_i = CZ_iC^*$.
\end{lemma}

\begin{proof}
    $U_i (C \ket{\vec 0}) = C Z_i \ket{\vec 0} = C \ket{\vec 0}$.  The $U_i$ are independent since the $Z_i$ are.
\end{proof}

We can rapidly extract a set of generators $P_i$ for this stabiliser subgroup using the techniques of the previous section.  Since $P_i$ and $U_i$ generate the same stabiliser subgroup, we have that $P_i = U^{\vec{\rho}_i}$ for some $\vec{\rho}_i \in \Z_2^n$.  Thus,
\begin{equation}\label{eqn:pi_eqn}
    P_i \,C \kz = U^{\vec{\rho}_i} C \kz = C Z^{\vec{\rho}_i} \kz = (-1)^{\vec{\rho}_i\cdot \vec{z}} C\kz.
\end{equation}

We can thus extract the vectors $\vec{\rho}_i$ by computing $P_i \,C \kz$ for $\vec{z}$ having Hamming weight 1.  

Note that Pauli-vector multiplication can be done more quickly than by using standard matrix-vector multiplication.  For a vector $\ket{v} = \sum_{\vec z \in \Z_2^n} v_{\vec z} \kz$, and a Pauli gate  $P = (-1)^{c} (-i)^{\vec{p} \cdot \vec{q}} Z^{\vec{p}} X^{\vec{q\,}}$, we have $$P\ket{v} = (-1)^{c + \vec p \cdot \vec q} (-i)^{\vec{p} \cdot \vec{q}} \sum_{\vec z \in \Z_2^n} (-1)^{\vec p \cdot \vec z} v_{\vec z + \vec q} \kz.$$  We can compute $P\ket{v}$  by instead reindexing the vector $\ket{v}$ and introducing both a global phase change and, for each index $\vec z \in \Z_2^n$, the sign change determined by $\vec p \cdot \vec z$.  This requires $O(Nn)$ time rather than the $O(N^2)$ time required for standard matrix-vector multiplication.

We know that the $\vec \rho_i$ are linearly independent since the $P_i$ are independent.  We may thus invert the matrix whose columns are  $\vec \rho_i$; we call the rows of this inverse $\vec \mu_i$.  It is a consequence of the following lemma that $U_i = P_i^{\vec \mu_i}$.

\begin{lemma}\label{mconjzlemma}
    Suppose that $M$, $U_i$ are matrices of size $N \times N$, with $M$ unitary, such that for all $\vec{z} \in \Z_2^n$ \begin{equation}\label{uimkz}U_i M \kz = (-1)^{z_i} M \kz.\end{equation}  Then $M$ conjugates the $Z_i$ to $U_i$; that is, $M Z_i M^* = U_i$.
\end{lemma}

\begin{proof}
    Multiplying \eqref{uimkz} by $\bra{\vec{z}_2} M^*$ we see that $\bra{\vec{z}_2} M^* U_i M \ket{\vec{z}_1} = (-1)^{z_i} \delta_{\vec{z}_1,\vec{z}_2}$ and thus $ M^* U_i M = Z_i$.
\end{proof}

Having found the $U_i$, we must now find the $V_i$. Our strategy will be to find Pauli gates $W_i$ of order 2 such that $U_i$ and $W_j$ anticommute if and only if $i=j$.  We will then correct these to get the $V_i$ by querying as few entries of $C$ as possible.

Constructing the $W_i$ is straightforward.  If $U_i \propto X^{\vec q_i}Z^{\vec p_i}$, define $U$ to be the $n \times 2n$ matrix over $\Z_2$ whose $i$-th row is $(\vec q_i, \vec p_i)$.  As these rows are linearly independent, $U^+ = U^T (UU^T)^{-1}$ is a right inverse: $UU^+ = \mathbb{I}$.  We may thus take $W_i$ to be $(-i)^{\vec \alpha_i \cdot \vec \beta_i} X^{\vec \beta_i}Z^{\vec \alpha_i}$ where $(\vec \alpha_i, \vec \beta_i)$ is the $i$-th column of $U^+$.

Each $V_i$ differs from $W_i$ by a product of the $U_j$ as a result of the following lemma.

\begin{lemma}\label{lem:V_and_W_relation}
Suppose the $n$ Pauli gates $U_i$ generate a stabiliser group, the Pauli gates $V_i$, $W_i$ are of order 2, and that $V_i$ commutes (or anticommutes) with $U_i$ if and only if  $W_i$ does.  Then $V_i = (-1)^{c_i}(-i)^{d_i} W_i U^{\vec v_i}$ for some $\vec v_i \in \Z_2^n$ and $c_i, d_i \in \Z_2$.
\end{lemma}

\begin{proof}
    Define the Pauli gates $T_i = i^{d_i}W_i V_i$ where $d_i$ is 0 if $W_i$ and $V_i$ commute and 1 if they anticommute.  They commute with every $U_j$: $T_i U_j = i^{d_i}W_i V_i U_j = i^{d_i}U_j W_i V_i = U_j T_i$.  Further, the $T_i$ are of order 2.  Therefore, by Definition \ref{stabsbgrpdefn}, for each $i$, either $T_i$ or $-T_i$ is a member of the stabiliser group generated by $U_i$.
\end{proof}

Therefore, our final task is to find the $V_i$ by determining the appropriate $\vec v_i \in \Z_2^n$ and $c_i, d_i \in \Z_2$.
\begin{align}\label{comparecols}
C\ket{\vec z + \vec e_i} &= C X_i \kz\\ 
&= V_i C \kz\\  
&= (-1)^{c_i}(-i)^{d_i} W_i U^{\vec v_i} C \kz\\ 
&= (-1)^{c_i}(-i)^{d_i} W_i C Z^{\vec v_i} \kz\\  
&= (-1)^{\vec v_i \cdot \vec z \,+\, c_i}(-i)^{d_i} W_i  \,\, C \kz\label{comparecols2}
\end{align}
where $\vec e_i$ is the $i$-th unit vector of $\Z_2^n$.

By comparing the relative phases of the columns of $C$ corresponding to $\vec z$ and $\vec z + \vec e_i$, we gain information about the required variables $\vec v_i, c_i, d_i$.  Choosing $\vec z = \vec 0$, we determine the $c_i, d_i$.  Choosing $\vec z$ to be of Hamming weight 1, we determine the components of the $\vec v_i$.  

To determine the relative phases of two columns, we need only check two nonzero entries: one from each column. To avoid potentially having to perform a high number of checks for columns that contain many zeros, we can first find an index in the support of each column. We use the fact that the support of $C\ket{\vec{z}}$ is the same as that of $W^{\vec{z}}C\ket{\vec{0}}$, as each $U_i$ either stabilises or antistabilises every column. We find the stabiliser state $C\ket{\vec{0}}$ in \Stwo\ form, which then allows us to use fast Pauli-vector multiplication to apply $W^{\vec{z}}$ and find an index in the support of $C\ket{\vec{z}}$. By subsequently applying $W_i$, we can find an index in the support of $C\ket{\vec{z} + \vec{e_i}}$.

\begin{algorithm}
{Extract the conjugate tuple of a Clifford gate $C$, i.e.\ \Cone\ $\to$ \Ctwo}
\label{c1toc2}
\item Compose Algorithms 1 and 4 to extract generators $P_i$ for the stabiliser group of $C\ket{\vec{0}}$.

\item For each $\vec z$ with Hamming weight 1, find a nonzero entry of $C\kz$. For each $i$, use these nonzero entries to determine $\vec{\rho}_i$ satisfying $P_i C \kz = (-1)^{\vec{\rho}_i \cdot \vec z} \kz$.  Note we do not have to verify that this equation holds; it is sufficient to check the action of $P_i^*$ on the nonzero entries of the columns to set $\vec{\rho}_i$.

\item Invert the $n \times n$ matrix over $\Z_2$ whose columns are $\vec{\rho}_i$ to find one with rows $\vec{\mu}_i$. Conclude that $U_i = P^{\vec \mu_i}$.

\item Compute the bitstrings specifying the $W_i$ (up to sign) by taking the pseudoinverse of the matrix whose rows are the bitstrings specifying the $U_i$.

\item Correct the $W_i$ to $V_i$ using Equations \eqref{comparecols} and \eqref{comparecols2} by comparing two nonzero entries of the $\vec z$-th and $\vec z + \vec e_i$-th columns of $C$ for $\vec z$ being $\vec 0$ or having Hamming weight 1.

\end{algorithm}

\subsection[Verifying a unitary matrix is a Clifford gate]{\Cone: Verifying a unitary matrix is a Clifford gate}

One can use Algorithm \ref{c1toc2} to verify that a given matrix $M$ is a Clifford gate by accepting only at the point that a conversion is made successfully.  Since we are not assuming that $M$ is a Clifford gate, we must add a few additional checks.

First, when we extract the $P_i$, we may not assume that $M\ket{0}$ is a stabiliser state; we explain how to verify this in Subsection \ref{sec:verif_stab_state}. We find the $U_i$ and $V_i$ as in Algorithm \ref{c1toc2}; the $U_i$ stabilise $M\ket{0}$ and $U_i V_j = (-1)^{\delta_{ij}} V_j U_i$ by construction.  Note that by Lemma \ref{lem:V_and_W_relation}, the $V_i$ will pairwise commute. We then make use of the following lemmata:

\begin{lemma}\label{lem:verifying_cliff}
    Let $M$ be an $N \times N$ unitary matrix. Suppose there exist Pauli gates $U_i, V_i \in \cP_n$ for  $i \in [n]$ such that:
    \begin{enumerate}[label=\roman*)]
        \item $MX_iM^* = V_i$,
        \item $U_i V_j = - V_j U_i$ if and only if $i = j$,
        \item $U_i M\ket{0} = M\ket{0}$.
    \end{enumerate}
    Then $MZ_iM^* = U_i$ and, hence, $M$ is a Clifford gate.
\end{lemma}
\begin{proof}
    $U_i M\kz \overset{i)}{=} U_i V^{\vec{z}} M\ket{\vec 0} \overset{ii)}{=} (-1)^{z_i} V^{\vec{z}} U_i M\ket{\vec 0} \overset{iii)}{=} (-1)^{z_i} V^{\vec{z}} M\ket{\vec 0} \overset{i)}{=} (-1)^{z_i}M\kz$. The result then follows from Lemma \ref{mconjzlemma}. 
\end{proof}

\begin{lemma}\label{mconjxlemma}
    Suppose that $M$, $V_i$ are matrices of size $N \times N$, with $M$ unitary, such that for all $\vec{z} \in \Z_2^n$ \begin{equation}\label{vimkz}V_i M \kz =  M \ket{\vec z + \vec e_i}.\end{equation}  Then $M$ conjugates the $X_i$ to $V_i$; that is, $M X_i M^* = V_i$.
\end{lemma}
\begin{proof}
    Similar to Lemma \ref{mconjzlemma}.
\end{proof}

By Lemma \ref{lem:verifying_cliff}, it is sufficient to check that $MX_iM^* = V_i$. By Lemma \ref{mconjxlemma}, it is sufficient to check that ${M\kz = V^{\vec{z}} M\ket{0}}$. To do this, we iterate through the Gray code \cite{Bitner1976gray}. 
Let $\vec{\alpha}_{i}$ be the $i$-th codeword.  We check that ${M\ket{\vec{\alpha}_{i}} = V^{\vec{\alpha}_{i} + \vec{\alpha}_{i-1}}M\ket{\vec{\alpha}_{i-1}}}$. The Gray code has the useful property that any two consecutive codewords differ only at a single bit, so that we only need to multiply our previously checked column by a single $V_i$ at each step.

\subsection[Conjugate tuple to a matrix]{\Ctwo\ $\to$ \Cone: Conjugate tuple to a matrix}

To rapidly construct the matrix of a Clifford gate (up to phase) given its conjugate tuple, we employ Theorem 3.9 of \cite{de2021efficient}.

\begin{theorem}\label{thm:matrix_from_conj_tuple}
The unitary $C$ that carries each $(Z_i,X_i)$ to $(U_i,V_i)$ under conjugation is given by: 
\begin{equation}C \kz = V^{\vec{z}} \ket{u_0} \end{equation}
where $\vec{z} \in \Z_2^n$ and $\ket{u_0} = C \ket{0}$ is a simultaneous eigenvector of the $U_1, \dots, U_n$ with eigenvalue 1.
\end{theorem}

The simultaneous eigenvector of the $U_1,\ldots,U_n$ with eigenvalue 1 can be quickly computed by combining Algorithms \ref{s3tos2} and \ref{s2tos1} above, giving the first column of the matrix. Then, to find the remaining columns, we iterate through the Gray code, using Theorem \ref{thm:matrix_from_conj_tuple}.  Every column will differ from the previous one by the application of a single $V_i$  that corresponds to the bit that was just flipped. Here, we again make use of the fast method for Pauli-vector multiplication.



\section{Complexity analyses}\label{sec:complexity_analysis}
We explicitly consider the complexity of Algorithms 1 through 5 as well as the other interconversion and verification algorithms based on them.  Recall that $n$ denotes the number of qubits and $N=2^n$.

\subsection{Stabiliser states}

\noindent\textbf{Algorithm \ref{s1tos2}: \Sone\ $\to$ \Stwo}
\begin{enumerate}
    \item Finding the nonzero amplitudes takes $O(N)$ time; applying a bitwise \texttt{XOR} between $\vec{z}_0$ and each binary label in the support then takes $O(2^k)$ operations, where $k$ is the dimension of the support.  We are modelling bitwise \texttt{XOR} as a constant-time operation as it is fully parallelisable.
    \item Extracting $\ell$ takes time $O(n)$.
    \item Extracting $Q$ takes time $O(n^2)$.
\end{enumerate}
The dominant step of the algorithm is Step 1 yielding a runtime of $O(N)$. \\

\noindent\textbf{Algorithm \ref{s2tos1}: \Stwo\ $\to$ \Sone}
\begin{enumerate}
    \item Initialising an empty state vector takes time $O(N)$.
    \item As noted in the main text, at each step of the Gray code, it takes time $O(n)$ to update $t_i$ and $\vec \beta_i$. Thus, this step runs in time $O(2^kn)$.
\end{enumerate}
In the case that $k$ is very small, the dominant step of the algorithm is Step 1. However, if $k = \Omega(n)$, Step 2 will dominate and Algorithm \ref{s2tos1} runs in time $O(N n)$. \\

\noindent\textbf{Algorithm \ref{s3tos2}: \Sthree\ $\to$ \Stwo}
\begin{enumerate}
    \item Gaussian elimination takes time $O(n^3)$.
    \item Finding a solution to the row reduced equation takes time $O(n)$.
    \item Computing $\ell$ takes time $O(k)$, where $k$ is the dimension of the affine subspace.
    \item Computing $Q$ takes time $O(k^2)$.
\end{enumerate}
Thus, Algorithm \ref{s3tos2} runs in time $O(n^3)$.\\

\noindent\textbf{Algorithm \ref{s2tos3}: \Stwo\ $\to$ \Sthree}
\begin{enumerate}
    \item Row reduction takes time $O(nk^2)$, where $k$ is the dimension of $V$. Finding a null basis for the null space then takes time $O((n-k)k)$.
    \item Computing $Q$ and $\ell$ can be done in parallel with Step 1, altering after each row reduction. Each update takes constant time (as it corresponds to swapping a row or adding two rows together) and thus this does not affect the complexity of Step 1.
    \item This takes time $O((n-k)n)$.
    \item This takes time $O(kn)$.
\end{enumerate}
In general, when $k = \Omega (n)$, Algorithm \ref{s2tos3} runs in time $O(n^3)$. \\

\noindent\textbf{\Sone\ $\to$ \Sthree:}  \\
Our algorithm is the composition of Algorithms \ref{s1tos2} and \ref{s2tos3}, and thus requires time $O(N)$.\\

\noindent\textbf{\Sthree\ $\to$ \Sone:} \\
Our algorithm is the composition of Algorithms \ref{s3tos2} and \ref{s2tos1}, and thus requires time $O(N n)$.\\


\noindent\textbf{Verifying \Sone:} \\
Our algorithm is the composition of Algorithm \ref{s1tos2} and Algorithm \ref{s2tos1}, followed by a consistency check between the original and reconsituted state vectors, and thus requires time $O(N)$. Note that to verify a vector, one must look at every element at least once, giving an $\Omega(N)$ lower bound.\\

\subsection{Clifford gates}

\noindent\textbf{Algorithm \ref{c1toc2}: \Cone\ $\to$ \Ctwo}
\begin{enumerate}
    \item The conversion \Sone\ to \Sthree\ takes $O(N)$ time as discussed above.
    \item In the worst case, we search for time $O(N)$ to find a nonzero element of a given column. Once we have found one for each column of Hamming weight 1, taking time $O(Nn)$, we can extract all the $\vec{\rho}_i$ vectors in time $O(n^2)$.
    \item It takes time $O(n^3)$ to find the pseudoinverse of the $U$.  In fact, the output of our \Sone\ to \Sthree\ conversion gave us the row-reduced version of $U$; by keeping track of the row operations when constructing the $U_i$ from the $P_i$, we can construct the pseudoinverse simultaneously. We then require time $O(n^2)$  to find each of the $W_i$.
    \item Comparing the relative phase of the columns corresponding to Hamming weight 1 and 2 takes time $O(n^3)$ as there are $n^2$ such columns and finding a nonzero element takes time $O(n)$. Then, finding the $V_i$ involves multiplying each $W_i$ by $O(n)$ Pauli gates, where each Pauli multiplication takes time $O(n)$.  This step thus also runs in time $O(n^3)$.
\end{enumerate}The algorithm is dominated by Step 2 and thus takes time $O(N n)$.\\

\noindent\textbf{\Ctwo\ $\to$ \Cone:}\\
The \Sthree\ $\to$ \Sone\ conversion to find the first column requires time $O(Nn)$. Computing the remaining columns using the Gray code takes time $O(N^2n)$, giving a total time of $O(N^2 n)$.\\

\noindent\textbf{Verifying \Cone:}\\
Compared to Algorithm \ref{c1toc2}, we must also verify that the first column of the unitary matrix is a stabiliser state and check the consistency of the remaining columns (using the Gray code). Verifying the first column takes time $O(N)$. At every step of the Gray code, the two consecutive codewords differ at a single bit, so we multiply our previously checked column by a single Pauli gate which takes time $O(Nn)$. We check every column, so the total time is $O(N^2n)$.

Note that to verify a matrix, one must look at every element at least once, giving an $\Omega(N^2)$ lower bound.

\section{Acknowledgments}
ND acknowledges support from the Canada Research Chair program, NSERC Discovery Grant RGPIN-2022-03103, the NSERC-European Commission project FoQaCiA, and the Faculty of Science of Simon Fraser University. WS acknowledges support from the EPSRC and Hitachi.  We thank Matthew Amy and Shane Mansfield for helpful comments.

\printbibliography
\newpage
\appendix

\section{Complexity comparisons to existing implementations}\label{sec:comparison_to_implementations}
In this appendix, we compare our algorithms to existing implementations in popular libraries: IBM's \texttt{Qiskit} \cite{Qiskit} package and Google's \Stim package due to Gidney \cite{gidney2021stim}.  We note that worst-case asymptotic complexity is a very coarse metric that does not fully capture the scope of an algorithm's advantage; it ignores differences in prefactors and lower order terms in their runtime costs. Establishing a difference is sufficient, however, to guarantee that one algorithm will outperform another for all sufficiently large inputs.\\

\noindent\textbf{\Sone\ $\to$ a succinct description of a stabiliser state and verifying \Sone:}\\
To our knowledge,  \texttt{Qiskit} does not implement a method for converting the amplitudes of a stabiliser state into a succinct description of that state. 

\Stim converts the amplitudes of a stabiliser state into a succinct representation in the form of a tableau using the \texttt{from\_state\_vector} function in the \texttt{Tableau} class. A key subroutine is the \texttt{stabiliser\_state\_vector\_to\_circuit} function in the file \texttt{circuit\_vs\_\linebreak amplitudes.cc}, which finds a Clifford circuit that transforms $\ket{\vec{0}}$ into the given stabliser state. On input $\ket{\psi}\in\C^N$ the algorithm runs as follows.
\begin{enumerate}
    \item By applying a series of $X$ gates, move the element with the largest amplitude to the first entry of the state vector.
    \item Find a nonzero element in the vector which is not the first entry; suppose it has index $\vec{k}\in \{0,1\}^n$. If no such entry exists, terminate the algorithm.
    \item Let $i\in[n]$ be the smallest index such that $\vec{k}_i=1$. Apply a series of CNOT gates between qubit $i$ and qubit $j$, for every $j\neq i$ such that $\vec{k}_j = 1$. Finally, apply a single-qubit Clifford gate to qubit $i$. If $\ket{\psi}$ was a stabiliser state, this will halve the size of the support of the state vector.  If this was not the case then output that the input was not a stabiliser state.
    \item Return to Step 2.
\end{enumerate}
Since the  support of the state vector is halved after every iteration, the algorithm will run Step 3 at most $k$ times, where $k$ is the dimension of the support of $\ket{\psi}$. Each iteration of Step 3 involves applying at most $\Omega(n)$ CNOT gates in the worse case. Each application of a CNOT gate to $\ket{\psi}$ requires $\Omega(N)$ steps. Thus, in the worst case, \texttt{Stim}'s algorithm runs in time $\Omega(Nn^2)$ and produces a circuit of size $\Omega(n^2)$. Note that it also verifies whether the input state is a stabiliser state. \\


\noindent\textbf{\Sthree\ $\to$ \Sone:}\\
We do not believe this is implemented directly in \texttt{Qiskit}. This is implemented in \texttt{Stim} by the function \texttt{to\_state\_vector} of the \texttt{Tableau} class.  The conversion works by:
\begin{enumerate}
    \item Generate a random vector in $\mathbb{C}^N$. It will have overlap with all stabiliser states with probability 1 in the ideal setting and probability very close to one in practice.
    \item For every Pauli gate $P_i$ in the check matrix, apply the projector $(P_i+\mathbb{I})/2$ to the current state vector. 
    \item Renormalise the state.
\end{enumerate}

Note that applying a projector $(\mathbb{I} + P)/2$ takes times $\Omega(Nn)$. This is done for each Pauli gate, giving a total runtime of $\Omega(Nn^2)$.  This algorithm is probabilistic and is not guaranteed to succeed on every run. \\


\noindent\textbf{Other stabiliser state conversions:}\\
We are not aware of any implementations of algorithms that convert to or from \Stwo\ or from \Sthree\ to \Stwo\ other than ours. \\

\noindent\textbf{\Cone\ $\to$ \Ctwo\ and verifying \Cone:}\\
Both \texttt{Qiskit} and \Stim implement \Cone\ $\to$ \Ctwo\ and additionally verify that the input matrix is a Clifford gate.

In \texttt{Qiskit}, the conversion is the \texttt{from\_matrix} method of the \texttt{Clifford} class. It is implemented via the brute force approach: conjugating each basic Pauli gate via matrix multiplication, which takes time $\Omega(N^3 n)$.

In \texttt{Stim}, the conversion is the \texttt{from\_unitary\_matrix} method of the \texttt{Tableau} class. Its implementation is more complex; on input matrix $C$ the algorithm runs as:
\begin{enumerate}
    \item Find a Clifford gate $U$ (and its circuit representation) such that $U\ket{\vec{0}} = C \ket{\vec{0}}$ (i.e.\ run the conversion described above on the first column of $C$).
    \item Calculate $M = U^\dag C$. If $C$ is a Clifford gate, then $M$ will be a Clifford gate that permutes the computational basis vectors with phases.
    \item Find a circuit implementing $M$. In this step,  $M$ is verified to be a Clifford gate which in turn verifies that $C$ is a Clifford gate.
    \item Concatenate the circuits for $U$ and $M$ to find a circuit representation of $C$.
    \item Convert this circuit representation into a stabiliser tableau.
\end{enumerate}
In particular, Step 2 involves multiplying $U^\dag$ and $M$. The circuit for $U$ has size $\Omega(n^2)$ in the worst case (see above), and multiplying $C$ by a one- or two-qubit gate takes time $\Omega(N^2)$. Thus Step 2, and hence \texttt{Stim}'s algorithm, runs in time $\Omega(N^2n^2)$.\\

\noindent\textbf{\Ctwo$\to$\Cone:}\\
Both \texttt{Qiskit} and \Stim implement \Ctwo\ $\to$ \Cone. In \texttt{Qiskit}, the conversion is the \linebreak\texttt{to\_operator} method of the \texttt{Clifford} class. The algorithm first decomposes the Clifford gate as a circuit, using the method of \cite{bravyi2021clifford}, and then finds the matrix of that circuit. If the circuit is size $s$, this compilation takes time $\Omega(sN^2)$. In the worst case, $s=\Omega(n^2)$ and thus \texttt{Qiskit}'s algorithm runs in time $\Omega(N^2n^2)$. In \texttt{Stim}, the conversion is the \texttt{to\_unitary\_matrix} method of the \texttt{Tableau} class. It takes $O(N^2n^2)$ time; as it relies on its implementation of the \Sthree\ to \Sone\ conversion, it is also probabilistic and not guaranteed to succeed on every run.
\newpage
\section{Timed benchmarks}\label{benchmarks}

We benchmarked our \texttt{C++} implementations of the ten algorithms for small numbers of qubits $n$ and, wherever possible, compared our implementations with existing ones from \texttt{Stim} \cite{gidney2021stim} and \texttt{Qiskit} \cite{Qiskit}. Uniformly random stabiliser states were generated using the techniques of \cite{randstab}; uniformly random Clifford gates were generated using the \texttt{qiskit.\linebreak quantum\_info.random\_clifford} function.
The benchmarking was done on a ThinkPad P16 Gen 1 PC running Ubuntu 22.04.5 LTS, with a 12th Gen Intel Core i7-12800HX CPU @ 2 GHz and 16 GiB of RAM. Our code was compiled using \texttt{gcc} version 11.4.0, and the wrapper code run using \texttt{Python} 3.10.12.

Our data is collected in Figure \ref{fig:benchmarking} below; note the log scale $y$-axis. Each test (a)--(j) was run with 1000 iterations. Wherever `succinct representation' appears in the title of a graph, we mean the standard way of storing a stabiliser state or Clifford gate in each implementation. In our code, stabiliser states are represented using the \texttt{Stabiliser\_State} class, which stores the \Stwo\ description, and Cliffords are represented using the \texttt{Clifford} class, which stores the \Ctwo\ description. \texttt{Stim} represents both stabiliser states and Cliffords using the \texttt{stim.Tableau} class, which is equivalent to the \Sthree\ and \Ctwo\ descriptions. \texttt{Qiskit} similarly represents Cliffords in the \Ctwo\ description using the \texttt{qiskit.quantum\_info.Clifford} class.

We found absolute timing advantages of some orders of magnitude for small numbers of qubits; our asymptotic complexity analyses guarantee arbitrarily large absolute advantages with sufficiently large $n$.

\vspace{0.1cm}
    


\begin{figure}[h]
    \caption{\centering Benchmarking results for our code vs.\ \texttt{Stim} and \texttt{Qiskit}. Each test (a)--(j) was run with 1000 iterations.}\label{fig:benchmarking}
    \centering
    \subfloat[\centering Testing a random mixture of valid stabiliser states and stabiliser states with one entry modified.]{\resizebox{0.49\textwidth}{!}{\input{plots/Testing_S_V.pgf}}}
    \subfloat[\centering Converting random stabiliser statevectors (amplitudes) into a succinct representation.]{\resizebox{0.49\textwidth}{!}{\input{plots/S_V_to_succinct_rep.pgf}}}\\
    \subfloat[\centering Performing the reverse conversion to (b) on random stabiliser states, starting with the succinct representation used by each library.]{\resizebox{0.49\textwidth}{!}{\input{plots/Succinct_rep_to_S_V.pgf}}}
    \subfloat[\centering Testing a random mixture of valid Clifford gate matrices and Clifford gate matrices with one entry modified.]{\resizebox{0.49\textwidth}{!}{\input{plots/Testing_C_U.pgf}}}
\end{figure}

\begin{figure}[h]
    \centering
    \subcaptionbox*{}{}
    \setcounter{subfigure}{4}
    \subfloat[\centering Converting random Clifford gate matrices into a succinct representation.]{\resizebox{0.49\textwidth}{!}{\input{plots/C_U_to_succinct_rep.pgf}}}
    \subfloat[\centering Performing the reverse conversion to (e) on random Clifford gate matrices, starting with the succinct representation used by each library.]{\resizebox{0.49\textwidth}{!}{\input{plots/Succinct_rep_to_C_U.pgf}}}\\
    \subfloat[\centering Converting random stabiliser statevectors into check matrix form.]{\resizebox{0.49\textwidth}{!}{\input{plots/S_V_to_S_P.pgf}}}
    \subfloat[\centering Converting random stabiliser check matrices into statevectors.]{\resizebox{0.49\textwidth}{!}{\input{plots/S_P_to_S_V.pgf}}}\\
    \subfloat[\centering Converting random stabiliser check matrices into our succinct representation, \Stwo.]{\resizebox{0.49\textwidth}{!}{\input{plots/S_P_to_succinct_rep.pgf}}}
    \subfloat[\centering Converting random stabiliser states from our succinct representation to check matrix form.]{\resizebox{0.49\textwidth}{!}{\input{plots/Succinct_rep_to_S_P.pgf}}}

\end{figure}

\end{document}